\documentclass[pdflatex,sn-mathphys-num]{sn-jnl}% Math and Physical Sciences Numbered Reference Style
\theoremstyle{thmstyleone}%
\theoremstyle{thmstyletwo}%

\theoremstyle{thmstylethree}%

\theoremstyle{definition}

\newcommand{\ket}[1]{\left|{#1}\right\rangle}

\usepackage{enumitem}
\newlist{steps}{enumerate}{1}
\setlist[steps, 1]{wide=0pt, leftmargin=\parindent, label=Step \arabic*:, font=\bfseries}
\usepackage{amssymb,amsmath,amsthm,amsfonts,amstext,amsbsy,amscd}
\usepackage{mathrsfs,mathtools,commath} % \norm ??? commath
\usepackage{color, soul} % text color, etc.
\usepackage{esvect,bm} % vector arrow
\usepackage{array} %????Table ??? ????????????
\usepackage{verbatim} % ??? ?? \begin{comment} \end{comment}
\usepackage{diagbox}
\usepackage{multirow}
\usepackage{amsmath}
\usepackage{tikz}
\usepackage{algorithm}
\usepackage{algpseudocode}
\usepackage{subcaption}

\usepackage{kotex}
\usepackage[table]{xcolor}
\usepackage{graphicx}%
\usepackage{multirow}%
\usepackage{amsmath,amssymb,amsfonts}%
\usepackage{amsthm}%
\usepackage{mathrsfs}%
\usepackage[title]{appendix}%
\usepackage{xcolor}%
\usepackage{textcomp}%
\usepackage{manyfoot}%
\usepackage{booktabs}%
\usepackage{algorithm}%
\usepackage{algorithmicx}%
\usepackage{algpseudocode}%
\usepackage{listings}%

\begin{document}
	
	\title[Article Title]{3D Stacked Surface-Code Architecture for Measurement-Free Fault-Tolerant Quantum Error Correction}
	
	%%=============================================================%%
	%% Prefix	-> \pfx{Dr}
	%% GivenName	-> \fnm{Joergen W.}
	%% Particle	-> \spfx{van der} -> surname prefix
	%% FamilyName	-> \sur{Ploeg}
	%% Suffix	-> \sfx{IV}
	%% NatureName	-> \tanm{Poet Laureate} -> Title after name
	%% Degrees	-> \dgr{MSc, PhD}
	%% \author*[1,2]{\pfx{Dr} \fnm{Joergen W.} \spfx{van der} \sur{Ploeg} \sfx{IV} \tanm{Poet Laureate} 
		%%                 \dgr{MSc, PhD}}\email{iauthor@gmail.com}
	%%=============================================================%%
	
	\author[1]{\fnm{GunSik} \sur{Min}}\email{mgs3351@korea.ac.kr}
	
	%\author[2,3]{\fnm{Second} \sur{Author}}\email{iiauthor@gmail.com}
	%\equalcont{These authors contributed equally to this work.}
	\author[2]{\fnm{IlKwon} \sur{Sohn}}\email{d2estiny@kisti.re.kr}
	
	%\author[2,3]{\fnm{Second} \sur{Author}}\email{iiauthor@gmail.com}
	%\equalcont{These authors contributed equally to this work.}

	\author*[1]{\fnm{Jun} \sur{Heo}}\email{junheo@korea.ac.kr}
	%\equalcont{These authors contributed equally to this work.}
	
	\affil*[1]{\orgdiv{School of Electrical Engineering}, \orgname{Korea University}, \orgaddress{\city{Seoul}, \postcode{02841}, \country{Republic of Korea}}}
    
	\affil[2]{\orgdiv{Quantum Network Research center}, \orgname{Korea Institute of Science and Technology Information}, \orgaddress{\city{Daejeon}, \postcode{34141}, \country{Republic of Korea}}}
	
	%\affil[2]{\orgdiv{Department}, \orgname{Organization}, \orgaddress{\street{Street}, \city{City}, \postcode{10587}, \state{State}, \country{Country}}}
	
	%\affil[3]{\orgdiv{Department}, \orgname{Organization}, \orgaddress{\street{Street}, \city{City}, \postcode{610101}, \state{State}, \country{Country}}}
	
	%%==================================%%
	%% sample for unstructured abstract %%
	%%==================================%%
	
	\abstract{
Mid-circuit measurements are a major bottleneck for superconducting quantum processors because they are slower and noisier than gates. Measurement-free quantum error correction (mfec) replaces repeated measurements and classical feed-forward by coherent quantum feedback, but existing mfec protocols suffer from severe connectivity overhead when mapped to planar surface-code architectures: transversal interactions between logical patches require SWAP chains of length $O(d)$ in the code distance, which increase depth and generate hook errors. This work introduces a 3D stacked surface-code architecture for measurement-free fault-tolerant quantum error correction that removes this connectivity bottleneck. Vertical transversal couplers between aligned surface-code patches enable coherent parity mapping and feedback with zero SWAP overhead, realizing constant-depth $O(1)$ inter-layer operations in $d$ while preserving local 2D stabilizer checks. A fault-tolerant mfec protocol for the surface code is constructed that suppresses hook errors under realistic noise. An analytical performance model shows that the 3D architecture overcomes the readout error floor and achieves logical error rates orders of magnitude below both standard measurement-based surface codes and 2D mfec variants in regimes with slow, noisy measurements, identifying 3D integration as a key enabler for scalable measurement-free fault tolerance.
}

	\keywords{Measurement-free quantum error correction, Surface code, 3D integration, Fault-tolerant quantum computing, Superconducting qubits}

	%%\pacs[JEL Classification]{D8, H51}
	
	%%\pacs[MSC Classification]{35A01, 65L10, 65L12, 65L20, 65L70}
	
	\maketitle

\section{Introduction}\label{sec:intro}

Among many quantum error-correcting architectures, the surface code has emerged as a leading candidate for large-scale fault-tolerant quantum computing on platforms with limited connectivity, including superconducting and neutral-atom devices~\cite{Sur12,XZZX,Stephens13}. While the threshold theorem guarantees scalable computation provided all primitive operations remain below a fault-tolerance threshold~\cite{Aharon97,Shor94,Shor96,Kitaev97,KITAEV2003,Aliferis08}, in practice the logical performance of near-term surface-code cycles is often limited by the properties of mid-circuit measurements~\cite{Chai2022}. On current superconducting hardware, single- and two-qubit gate errors have reached the $10^{-3}$--$10^{-4}$ regime, whereas readout remains slower and typically an order of magnitude noisier, with additional latency from classical processing~\cite{suppressing_2023,Graham23,Lis23,Blinov02}. This asymmetry can impose a practical ``readout floor'' in measurement-based surface-code (SM-SC) cycles: even as coherent gates improve, repeated measurement windows introduce long idles and extra noise channels that dominate the logical error budget~\cite{Aliferis08,suppressing_2023,Graham23,Lis23}.

Measurement-free quantum error correction (MFEC) addresses this bottleneck by replacing projective stabilizer measurements and classical feed-forward with coherent quantum feedback~\cite{Paz10,Crow16,Ercan18,Perlin2023FTMF}. In MFEC schemes, stabilizer information is extracted unitarily into auxiliary registers, processed coherently, and then fed back to the data block without collapsing the state~\cite{Paz10,Crow16,Ercan18,Perlin2023FTMF}. Recent circuit-level constructions have demonstrated fully fault-tolerant MFEC protocols for CSS codes using a logical auxiliary block together with unencoded ancillas, and have developed resource-optimized variants and measurement-free non-Clifford primitives tailored to specific platforms~\cite{Heussen2024MFEC,Veroni2024OptimizedMFEC,ButtMF,GiudiceMF,Veroni2025UniversalMFEC,Perlin2023FTMF}. These results establish MFEC as an engineering-relevant alternative when mid-circuit readout is the dominant limitation.

However, embedding MFEC into realistic \emph{planar} surface-code architectures reveals a severe connectivity bottleneck. A generic MFEC round for a logical surface-code qubit requires transversal interactions between a data patch and one or more logical auxiliary patches. On a strictly 2D nearest-neighbor layout, corresponding physical qubits in different patches are separated by $O(d)$ lattice spacings for code distance $d$, so transversal CNOT layers must be synthesized using SWAP chains of length $O(d)$. These SWAP networks inflate depth and idle time, and can introduce hook-error mechanisms in which a single fault spreads into an uncorrectable multi-qubit error within a block, jeopardizing the usual distance-$d$ fault-tolerance condition~\cite{Aliferis08}. Consequently, the measurement advantage of MFEC can be negated by routing overhead on 2D hardware as $d$ increases.

In this work we show that three-dimensional (3D) integration can resolve this bottleneck by providing \emph{native transversal connectivity} between stacked surface-code patches. Motivated by rapid progress in 3D superconducting integration---including flip-chip bonding, multilayer packaging, and superconducting through-silicon vias (TSVs) that enable vertical coupling while maintaining long coherence times---we treat 3D connectivity as a logical resource rather than merely a wiring solution~\cite{rosenberg_3d_2017,Mallek21,akhtar_high-fidelity_2023,YostTSV,Norris3D}. We propose a 3D stacked surface-code architecture in which three rotated surface-code layers encoding $\ket{0}_L$, $\ket{\psi}_L$, and $\ket{+}_L$ are vertically aligned and coupled column-wise. Vertical transversal couplers enable inter-layer parity mapping and feedback in constant depth $O(1)$ with respect to $d$, eliminating the $O(d)$ SWAP overhead while preserving local 2D stabilizer checks within each layer~\cite{Sur12,Stephens13}.

Building on the MFEC framework of Heu{\ss}en \emph{et al.}~\cite{Heussen2024MFEC}, we construct a fault-tolerant MFEC protocol for rotated surface-code patches in this 3D architecture and analyze its performance under a decoherence-based noise model calibrated to superconducting timescales~\cite{suppressing_2023,Graham23,Lis23}. We compare three settings on equal footing: standard measurement-based surface codes (SM-SC), 2D MFEC surface codes with SWAP-mediated transversal operations (2D MFEC-SC), and the proposed 3D stacked MFEC surface code (3D MFEC-SC). Our analysis identifies regimes with slow, noisy measurements in which 3D MFEC-SC overcomes the readout floor and achieves substantially lower logical error per cycle than both SM-SC and 2D MFEC-SC, highlighting 3D integration as a key enabler for scalable measurement-free fault tolerance~\cite{Aliferis08,suppressing_2023,Graham23,Lis23}.

The remainder of the paper is organized as follows. Sec.~\ref{sec2} fixes notation and reviews the minimal MFEC and rotated surface-code background needed for our constructions~\cite{Shor96,Steane96,Sur12}. Sec.~\ref{Formulation} then introduces the architectural bottleneck of planar nearest-neighbor layouts for MFEC---namely the $O(d)$ routing depth required for transversal coupling---and presents the proposed 3D stacked surface-code architecture together with a high-level MFEC cycle overview enabled by vertical couplers~\cite{Aliferis08,rosenberg_3d_2017,Mallek21,akhtar_high-fidelity_2023,YostTSV,Norris3D}. Detailed 2D circuit schedules and SWAP-based transversal routing constructions are deferred to Appendix~\ref{app:2d_mfec_details}~\cite{Heussen2024MFEC,Veroni2024OptimizedMFEC}. Sec.~\ref{sec:analysis} provides resource and logical performance comparisons under a unified decoherence-based noise model~\cite{suppressing_2023,Graham23,Lis23}. Sec.~\ref{Conclusion} concludes with implications for 3D-integrated, measurement-free fault-tolerant quantum processors~\cite{rosenberg_3d_2017,Mallek21,akhtar_high-fidelity_2023,YostTSV,Norris3D}.

%%%%%%%%%%%%%%%%%%%%%%%%%%%%%%%%%%%%%%%%%%%%%%%%%%%%%%%%%%

\section{Preliminaries}\label{sec2}

In this section we fix notation and recall only the minimal background needed to describe our measurement-free surface-code constructions. We assume familiarity with stabilizer quantum error correction~\cite{Shor96,Steane96,Sur12}, and focus on (i) the abstract structure of measurement-free QEC (MFEC) for CSS codes and (ii) the rotated surface-code instance that we use throughout. More pedagogical examples and circuit-level details are deferred to the Appendices.

\subsection{Measurement-free QEC for CSS codes}\label{subsec:mf_css}

Conventional stabilizer-based QEC relies on repeated projective measurements of stabilizer generators followed by classical decoding and conditional feedback~\cite{Shor96,Steane96,Sur12,Aliferis08}. In many platforms, however, mid-circuit measurements are substantially slower and noisier than entangling gates~\cite{Aliferis08,suppressing_2023,Graham23,Lis23}. MFEC replaces this measurement-and-feed-forward loop by a fully coherent protocol: stabilizer information is extracted into auxiliary registers via unitary operations and then coherently fed back to the data block without collapsing the state~\cite{Paz10,Crow16,Ercan18,Perlin2023FTMF}.

For an $[n,k,d]$ CSS code with stabilizer generators $\{S_1,\dots,S_{n-k}\}$, a generic MFEC step proceeds in two stages. First, syndrome information is coherently mapped from the data block onto an ancilla register $a_1,\dots,a_{n-k}$ via a stabilizer-extraction unitary
\begin{equation}
    U_{\mathrm{ext}}
    = \prod_{j=1}^{n-k} \mathrm{CNOT}\bigl(S_j \rightarrow a_j\bigr),
    \label{eq:generic_ext}
\end{equation}
where $\mathrm{CNOT}(S_j\rightarrow a_j)$ denotes a product of CNOT gates that maps the eigenvalue of $S_j$ onto the ancilla qubit $a_j$ (or, more generally, an entangling gate that performs the same mapping). Second, a coherent correction unitary
\begin{equation}
    U_{\mathrm{corr}}
    = \sum_{s\in\{0,1\}^{n-k}} P_s \otimes C_s,
    \label{eq:generic_corr}
\end{equation}
is applied, where $P_s$ projects the ancilla register onto a syndrome state $\ket{s_1\dots s_{n-k}}$ and $C_s$ is a Pauli correction acting on the data block. In a fault-tolerant MFEC design, the multi-controlled operations $C_s$ and the extraction network $U_{\mathrm{ext}}$ are arranged so that any single physical fault during the entire MFEC step produces at most one effective physical error per code block at the end of the cycle~\cite{Perlin2023FTMF,Heussen2024MFEC}. This ensures the usual distance-$d$ fault-tolerance condition and yields logical failure rates with the expected $O(p^2)$ scaling in the physical error rate $p$ for small codes~\cite{Perlin2023FTMF,Heussen2024MFEC,Veroni2024OptimizedMFEC}.

Heu{\ss}en \emph{et al.}\ formulated a systematic MFEC scheme for CSS codes based on three logical registers: a data block, a logical auxiliary block of the same code, and a register of unencoded ancillas~\cite{Heussen2024MFEC}. For $X$-error correction, a transversal CNOT is applied from the data block to the auxiliary block, coherently mapping $X$ errors while preserving the logical state; the auxiliary block is then processed and used to drive a coherent feedback operation on the data block. An analogous sequence corrects $Z$ errors. Subsequent work by Veroni, M{\"u}ller, Giudice and collaborators optimized these ideas for neutral-atom architectures and extended them towards universal MFEC, combining redundant syndrome extraction, flag qubits, and resource-efficient Toffoli/CCZ gadgets~\cite{Veroni2024OptimizedMFEC,ButtMF,GiudiceMF,Veroni2025UniversalMFEC}. In this paper we adopt the abstract MFEC structure of Eqs.~\eqref{eq:generic_ext}–\eqref{eq:generic_corr} and the three-block architecture of Ref.~~\cite{Heussen2024MFEC}, but specialize them to surface-code patches under realistic connectivity constraints.

A fully worked-out MFEC example for a small code (e.g.\ the three-qubit bit-flip code) is useful pedagogically but not essential for the main development here; a representative construction is therefore provided in Appendix~\ref{app:bitflip} rather than in the main text.

\subsection{MFEC for rotated surface codes and prior constructions}\label{subsec:rotated_mfec}

We now fix the specific surface-code instance used throughout the paper and briefly summarize prior MFEC surface-code constructions. We consider the rotated surface code with distance $d$, whose data qubits live on the vertices of an $L\times L$ rotated square lattice with $L=d$, and whose stabilizers are weight-2 and weight-4 plaquette operators~\cite{Sur12,Stephens13}. For the distance-$3$ case ($L=3$), a convenient choice of stabilizer generators acting on nine data qubits $\{1,\dots,9\}$ is
\begin{align}
    S_1^Z &= Z_1 Z_2 Z_5 Z_6,\quad
    S_2^Z = Z_4 Z_5 Z_8 Z_9, \nonumber\\
    S_3^Z &= Z_3 Z_4,\qquad\ \ 
    S_4^Z = Z_6 Z_7, \nonumber\\
    S_1^X &= X_2 X_3 X_4 X_5,\quad
    S_2^X = X_5 X_6 X_7 X_8, \nonumber\\
    S_3^X &= X_1 X_2,\qquad\ \ 
    S_4^X = X_8 X_9,
    \label{eq:rotated_stabs}
\end{align}
so that four $Z$-type and four $X$-type syndromes can be extracted in each round. Higher-distance rotated codes follow the same pattern, with $O(d^2)$ data qubits and $O(d^2)$ stabilizer checks~\cite{Sur12,Stephens13}.

Explicit MFEC circuits for the distance-3 rotated surface code have been constructed in Ref.~\cite{Veroni2024OptimizedMFEC} using the general scheme of Ref.~\cite{Heussen2024MFEC}. In an idealized connectivity setting, those circuits satisfy the single-fault tolerance condition at the circuit level and achieve logical failure rates consistent with $O(p^2)$ scaling. In particular, they demonstrate that MFEC can be implemented directly on small surface-code patches without mid-circuit measurements, using coherent syndrome extraction onto unencoded ancillas followed by multi-controlled corrections acting on the data block and a logical auxiliary block.

In this work we do not re-derive those MFEC circuits in detail; instead, we use the rotated surface code of Eq.~\eqref{eq:rotated_stabs} as a concrete testbed and adopt the same logical structure of the MFEC step (data block + logical auxiliary block + unencoded ancillas). Our focus in Sec.~\ref{Formulation} is on how this MFEC step can be embedded into physically realistic architectures with constrained connectivity: first, in a strictly 2D nearest-neighbor (NN) layout using SWAP-mediated transversal gates, and second, in a 3D stacked layout with native vertical couplers that remove the SWAP overhead. The noise model and parameter regimes used to compare these architectures—standard measurement-based surface code (SM-SC), 2D MFEC-SC, and 3D MFEC-SC—are introduced in Sec.~\ref{sec:analysis}, with detailed resource counts summarized in Appendix~\ref{app:gatecounts}.

%%%%%%%%%%%%%%%%%%%%%%%%%%%%%%%%%%%%%%%%%%%%%%%%%%%%%%%%%%
\section{3D Stacked Surface-Code Architecture and MFEC Protocol}\label{Formulation}

This section introduces a measurement-free surface-code implementation that is fault-tolerant at the circuit level and compatible with realistic hardware constraints. The core message is architectural: while MFEC removes mid-circuit measurements and classical feed-forward~\cite{Perlin2023FTMF,Heussen2024MFEC,Veroni2024OptimizedMFEC,Veroni2025UniversalMFEC}, a planar nearest-neighbor layout makes the required transversal interactions between logical patches expensive due to routing. We therefore propose a 3D stacked layout with native vertical couplers, enabling transversal coupling in constant depth with respect to the code distance $d$.

\subsection{Planar nearest-neighbor limitation: routing overhead for transversal coupling}\label{subsec:planar_limitations}

Fault-tolerant MFEC protocols for CSS codes typically rely on coherent syndrome extraction and coherent feedback mediated by an auxiliary logical block~\cite{Perlin2023FTMF,Heussen2024MFEC}. When specialized to surface-code patches, this structure requires \emph{transversal interactions} between a data patch and one (or more) logical ancilla patches~\cite{Veroni2024OptimizedMFEC}. In an idealized connectivity setting, such transversal CNOT layers can be executed in constant depth. On a strictly planar 2D nearest-neighbor (NN) architecture, however, corresponding physical qubits across distinct patches are separated by $O(d)$ lattice steps for code distance $d$. Implementing a transversal CNOT therefore requires routing (e.g., SWAP chains) of depth $O(d)$ per coupling layer, which increases the number of two-qubit-gate and idle locations per MFEC cycle and can exacerbate correlated-error mechanisms that must be carefully suppressed to maintain distance-$d$ fault tolerance~\cite{Aliferis08}. More broadly, as systems scale, the ability to execute gates in parallel is itself constrained by unwanted couplings and noise in realistic qubit arrays; for example, simulations of donor/quantum-dot flip-flop qubit arrays explicitly quantify how idle errors and simultaneous (parallel) gating degrade gate fidelity in small 1D/2D layouts~\cite{DeMichielis2024FFParallel}.

\begin{figure}[t]
    \centering
    \includegraphics[width=0.7\textwidth]{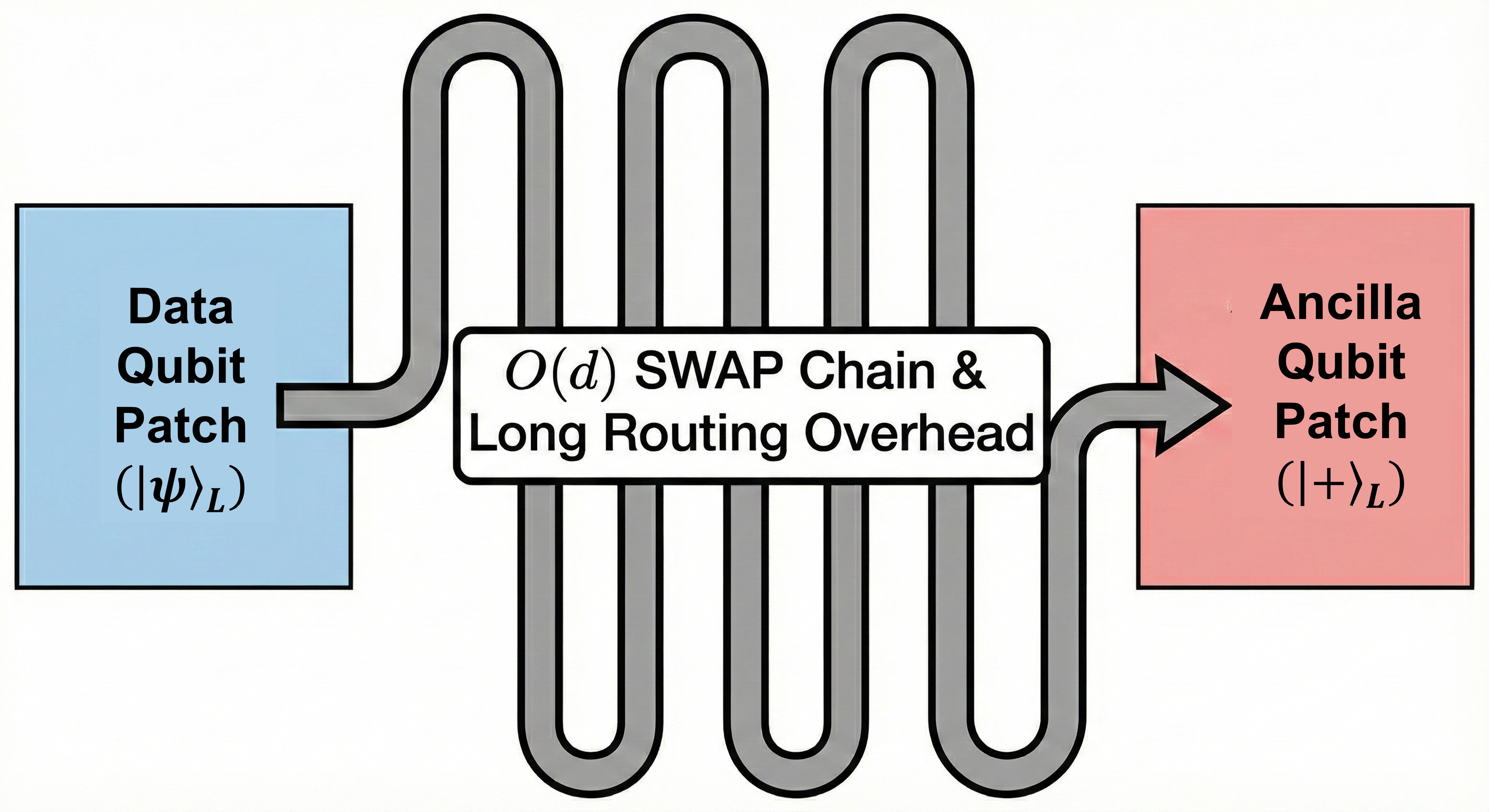}
    \caption{\textbf{Routing overhead for planar MFEC surface codes.}
    On a 2D nearest-neighbor layout, transversal coupling between a data patch and a logical ancilla patch requires SWAP-based routing with depth $O(d)$ for code distance $d$, increasing circuit depth and idling overhead.}
    \label{fig:2d_routing_cartoon}
\end{figure}

Fig.~\ref{fig:2d_routing_cartoon} summarizes this limitation at the architectural level: the transversal coupling needed for MFEC becomes a long routing problem in 2D. This is particularly problematic in the regime that motivates MFEC on superconducting hardware, where mid-circuit measurements are substantially slower and noisier than coherent gates~\cite{Aliferis08,suppressing_2023,Graham23,Lis23}. Although MFEC avoids measurement windows, a planar realization can lose much of the benefit if SWAP routing dominates the depth and idling overhead. Circuit-level 2D realizations and SWAP-based transversal schedules are therefore deferred to Appendix~\ref{app:2d_mfec_details}, and the main text focuses on an architecture that removes the routing bottleneck.

\subsection{3D stacked architecture and MFEC cycle overview}\label{subsec:3d_arch_and_cycle}

To eliminate the planar routing overhead, we propose a \emph{3D stacked surface-code architecture} enabled by vertical integration technologies such as flip-chip bonding, multilayer packaging, and superconducting through-silicon vias (TSVs)~\cite{rosenberg_3d_2017,Mallek21,akhtar_high-fidelity_2023,YostTSV,Norris3D}. The architecture stacks three distance-$d$ rotated surface-code patches and provides native couplers between vertically aligned physical qubits. This converts the transversal coupling required by MFEC into a set of parallel inter-layer two-qubit gates, eliminating SWAP routing between patches.

\begin{figure}[t]
    \centering
    \includegraphics[width=0.88\textwidth]{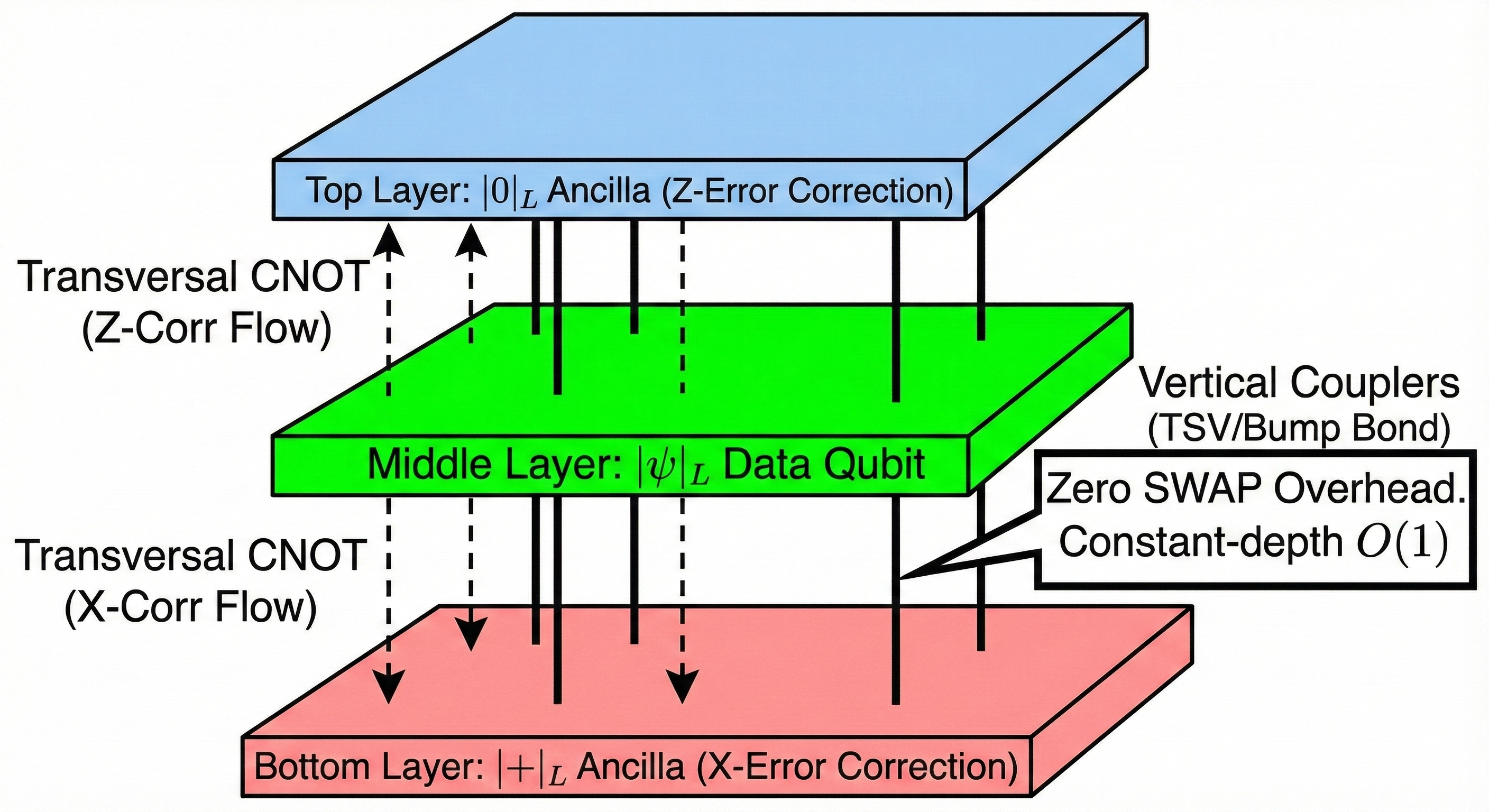}
    \caption{\textbf{Proposed 3D stacked surface-code architecture.}
    Three distance-$d$ rotated surface-code patches are stacked vertically.
    The middle layer hosts the data block $\ket{\psi}_L$, while the top and bottom layers are logical ancillas prepared in $\ket{0}_L$ and $\ket{+}_L$ for $Z$- and $X$-error correction, respectively.
    Vertical couplers enable transversal inter-layer coupling in constant depth with respect to $d$, removing the $O(d)$ SWAP-routing overhead required in planar 2D NN layouts.}
    \label{fig:3d_arch}
\end{figure}

We stack three rotated patches (linear size $L=d$): a top logical ancilla prepared in $\ket{0}_L$, a middle data patch encoding $\ket{\psi}_L$, and a bottom logical ancilla prepared in $\ket{+}_L$ (Fig.~\ref{fig:3d_arch}). The data patch is placed in the middle so that it can couple transversally to both ancilla layers. Vertical couplers connect aligned physical qubits across layers, enabling transversal CNOT layers between the data and either ancilla patch by applying parallel vertical two-qubit gates on all columns.

Within each layer, stabilizer checks and local operations proceed as in conventional surface-code schedules: all stabilizer generators involve only nearest neighbors in the $x$--$y$ plane~\cite{Sur12,Stephens13}. The role of the third dimension is targeted: it provides \emph{inter-patch} connectivity without sacrificing the favorable 2D locality of the surface code.

On this architecture, we implement a measurement-free QEC cycle following fault-tolerant MFEC principles for CSS codes~\cite{Perlin2023FTMF,Heussen2024MFEC}. Each cycle consists of two symmetric stages:

\begin{enumerate}[leftmargin=1.2em]
\item \textbf{$X$-error correction using $Z$-type checks (via $\ket{+}_L$ ancilla).}
A stabilizer-extraction network coherently maps $Z$-type syndrome information into an ancilla register. The resulting syndrome is processed coherently and used to drive conditional $X$ corrections on the data block, mediated through the logical $\ket{+}_L$ layer so that single faults do not fan out into uncorrectable multi-qubit data errors~\cite{Aliferis08,Perlin2023FTMF,Heussen2024MFEC}.

\item \textbf{$Z$-error correction using $X$-type checks (via $\ket{0}_L$ ancilla).}
Analogously, $X$-type syndrome information is coherently extracted and used to apply conditional $Z$ corrections on the data block, mediated through the logical $\ket{0}_L$ layer~\cite{Perlin2023FTMF,Heussen2024MFEC}.
\end{enumerate}

The crucial architectural difference from planar MFEC is that the \emph{inter-block transversal coupling} required to transfer parity/syndrome information between the data and logical ancilla layers is implemented by parallel vertical couplers, with depth independent of $d$ up to a hardware-dependent constant. This removes the dominant $O(d)$ routing overhead (Fig.~\ref{fig:2d_routing_cartoon}), reduces idling time, and lowers the effective number of fault locations per EC cycle compared to planar SWAP-based realizations.

We emphasize that the 3D stack does not make the \emph{entire} MFEC cycle independent of $d$: the number of stabilizers and intra-layer extraction workload still scale as $O(d^2)$~\cite{Sur12,Stephens13}. What becomes constant-depth (in $d$) is the inter-layer transversal coupling and associated parity mapping/feedback that would otherwise require $O(d)$ SWAP chains in planar NN layouts. In Sec.~\ref{sec:analysis} we quantify how this removal of routing overhead reduces effective fault-location counts and improves logical performance in regimes with slow and noisy measurements~\cite{suppressing_2023,Graham23,Lis23}.

%%%%%%%%%%%%%%%%%%%%%%%%%%%%%%%%%%%%%%%%%%%%%%%%%%%%%%%%%%%%%%%%%%%%%%%%
%%%%%%%%%%%%%%%%%%%%%%%%%%%%%%%%%%%%%%%%%%%%%%%%%%%%%%%%%%%%%%%%%%%%%%%%
\section{Performance analysis}\label{sec:analysis}

In this section, we compare three architectures:
(i) a conventional measurement-based surface code (SM-SC) with repeated stabilizer measurements,
(ii) a 2D measurement-free surface code with SWAP-based routing (2D MFEC-SC),
and (iii) the proposed 3D stacked measurement-free surface code (3D MFEC-SC).
We first summarize the noise model and parameter choices, then discuss resource requirements, and finally present logical error rates obtained from a combination of circuit-level simulation (SM-SC) and coarse-grained Monte Carlo sampling (MFEC-SC).

\subsection{Noise model and physically motivated parameters}\label{subsec:noise_model}

We use a simple decoherence–based model that separates
(i) gate errors,
(ii) idling errors during coherent operations,
and (iii) idling/measurement errors during readout windows.
Decoherence over a time interval $t$ is modeled via effective Pauli noise
\begin{equation}
    p_{\mathrm{decoh}}(t;T_1,T_2)
    = 1 - (1 - p_{\mathrm{relax}})(1 - p_{\mathrm{deph}}),
\end{equation}
with
\begin{equation}
    p_{\mathrm{relax}} = 1 - e^{-t/T_1}, \qquad
    p_{\mathrm{deph}}  = 1 - e^{-t/T_2},
\end{equation}
where $T_1$ and $T_2$ denote relaxation and dephasing times, respectively.
This mapping is not intended as a microscopic model but as a consistent way to translate $(T_1,T_2)$ and operation times into effective error probabilities that can be applied uniformly across all three architectures.

We distinguish the following time scales:
single-qubit gate time $t_{1\mathrm{q}}$, two-qubit gate time $t_{2\mathrm{q}}$, a typical idle slot during coherent operations $t_{\mathrm{idle,op}}$, and a measurement time $t_{\mathrm{m}}$.
Representative base values, inspired by current superconducting platforms, are listed in Table~\ref{tab:params}.
Implementation errors (e.g.\ control infidelity) are added on top of the decoherence contribution as small constants $p^{\mathrm{impl}}$, and the total probability is clipped to unity if necessary.

\begin{table}[htb]\centering
\caption{Representative hardware parameters used in the performance analysis. The overall noise level can be varied by rescaling $(T_1,T_2)$ by a factor $s$ (see text).}
\label{tab:params}
\begin{tabular*}{\textwidth}{@{\extracolsep{\fill}} lcc @{}}
\toprule
Quantity & Symbol & Value (base) \\
\midrule
Relaxation time           & $T_1^{\mathrm{(base)}}$         & $200~\mu\mathrm{s}$ \\
Dephasing time            & $T_2^{\mathrm{(base)}}$         & $150~\mu\mathrm{s}$ \\
Single-qubit gate duration& $t_{1\mathrm{q}}$               & $20~\mathrm{ns}$ \\
Two-qubit gate duration   & $t_{2\mathrm{q}}$               & $40~\mathrm{ns}$ \\
Idle slot during gates    & $t_{\mathrm{idle,op}}$          & $40~\mathrm{ns}$ \\
Measurement time          & $t_{\mathrm{m}}$                & variable \\
\midrule
Two-qubit impl.\ error    & $p_{2\mathrm{q}}^{\mathrm{impl}}$   & $2\times10^{-4}$ \\
Measurement impl.\ error  & $p_{\mathrm{meas}}^{\mathrm{impl}}$ & $5\times10^{-3}$ \\
\bottomrule
\end{tabular*}
\end{table}

For the SM-SC, we generate noisy rotated-surface-code memory circuits using Stim
with one round of syndrome extraction and decode them with MWPM (PyMatching). While we use fixed weights from a decoherence-motivated noise mapping for the SM-SC simulations, data-driven adaptive weight estimation can mitigate time-dependent drift of the effective noise rates without requiring an explicit error model~\cite{Spitz2018AdaptiveWeight}.

In these circuits, we assign
\begin{itemize}
    \item a depolarizing error with probability
    $p_{2\mathrm{q}}$ after each two-qubit gate,
    \item an idle error with probability $p_{\mathrm{idle,op}}$ on data qubits between rounds,
    \item and a flip error with probability $p_{\mathrm{meas}}$ before/after each measurement and reset.
\end{itemize}
These are obtained from the decoherence model as
\begin{align}
    p_{2\mathrm{q}}      &= p_{\mathrm{decoh}}(t_{2\mathrm{q}};T_1,T_2) + p_{2\mathrm{q}}^{\mathrm{impl}}, \\
    p_{\mathrm{idle,op}} &= p_{\mathrm{decoh}}(t_{\mathrm{idle,op}};T_1,T_2),\\
    p_{\mathrm{meas}}    &= p_{\mathrm{decoh}}(t_{\mathrm{m}};T_1,T_2) + p_{\mathrm{meas}}^{\mathrm{impl}}.
\end{align}

To capture the impact of slow, noisy measurements, we introduce the decoherence ratio
\begin{equation}
    R_{\mathrm{decoh}}
    = \frac{p_{\mathrm{decoh}}(t_{\mathrm{m}};T_1,T_2)}
           {p_{\mathrm{decoh}}(t_{\mathrm{idle,op}};T_1,T_2)},
\end{equation}
which compares decoherence accumulated during the measurement window to decoherence during a typical idle slot in a gate layer.
By varying $t_{\mathrm{m}}$ at fixed $(T_1,T_2)$, we scan $R_{\mathrm{decoh}}$ over experimentally relevant values.

Finally, to explore different overall noise levels while keeping relative time scales fixed, we rescale
\begin{equation}
    T_1 = \frac{T_1^{\mathrm{(base)}}}{s}, \qquad
    T_2 = \frac{T_2^{\mathrm{(base)}}}{s},
\end{equation}
with a “noise scale” $s$.
Larger $s$ corresponds to shorter coherence times and hence larger physical error probabilities for all architectures.

Details of how these parameters are mapped to effective fault locations for the MFEC architectures (including $N_{\mathrm{gate}}$ and $N_{\mathrm{idle}}$ per EC cycle) are given in Appendix~\ref{app:gatecounts};
here we only use these counts as effective coefficients in the logical error estimates.

\subsection{Resource estimates for the three architectures}\label{subsec:resources}

We next summarize the qubit overhead and CNOT-depth per error-correction (EC) cycle for the three architectures.
For the SM-SC, we consider a rotated surface code with the usual “one round per cycle” schedule for each distance $d$.
For the MFEC architectures, we base our counts on explicit distance-$3$ and distance-$5$ constructions and use their scaling in $d$ (Appendix~\ref{app:gatecounts}) to extrapolate.

Table~\ref{tab:resources} lists the qubit count and approximate CNOT-depth per EC cycle for representative distances.
The 2D and 3D MFEC architectures use the same number of physical qubits for a given $d$ (three stacked patches versus three planar patches), but differ significantly in depth: in 2D, transversal interactions between patches require SWAP chains of length $O(d)$, whereas in 3D they are implemented by single-layer vertical CNOTs.

\begin{table}[htb!]\centering\small
\setlength{\tabcolsep}{4pt}
\caption{Resource estimates for a single error-correction (EC) cycle in the three architectures:
3D measurement-free surface code (3D MFEC-SC),
2D measurement-free surface code (2D MFEC-SC),
and standard measurement-based surface code (SM-SC).
Depth values are approximate CNOT-depths.
For the 3D MFEC-SC, $X$- and $Z$-error correction rounds are pipelined, so the total depth is close to that of a single round (Appendix~\ref{app:gatecounts}).}
\label{tab:resources}
\begin{tabular}{@{}|c|c|c|c|c|@{}}
\hline
\textbf{Resource} & \textbf{Distance} & \textbf{3D MFEC-SC} & \textbf{2D MFEC-SC} & \textbf{SM-SC} \\
\hline
\multirow{3}{*}{\textbf{Qubits}} 
  & $d=3$ & $35$ & $35$ & $17$ \\
  & $d=5$ & $99$ & $99$ & $49$ \\
  & $d$   & $4d^2-1$ & $4d^2-1$ & $2d^2-1$ \\
\hline
\multirow{3}{*}{\textbf{CNOT depth / EC cycle}} 
  & $d=3$ & $59$  & $203$ & $\sim 8$ \\
  & $d=5$ & $158$ & $558$ & $\sim 8$ \\
  & $d$   & $O(7d^2)$ & $O(23d^2)$ & $O(1)$ \\
\hline
\end{tabular}
\end{table}

In the analytical model, these depths translate into effective numbers of faulty locations per EC cycle.
For a distance-$d$ code that can correct $t=(d-1)/2$ arbitrary physical errors per block, the leading logical error per cycle can be written schematically as
\begin{equation}
    p_L \;\sim\; \sum_{j=0}^{t+1}
    c_j\, p_{\mathrm{gate}}^{\,j}\, p_{\mathrm{idle}}^{\,t+1-j},
\end{equation}
where the coefficients $c_j$ scale with the number of gate and idle locations in the cycle.
The main architectural difference between 2D and 3D MFEC appears here as a difference in these coefficients: the 3D stack has significantly fewer effective locations because SWAP chains are removed and $X/Z$ rounds are pipelined.
A more detailed derivation and the explicit numbers used in our Monte Carlo sampling are given in Appendix~\ref{app:gatecounts}.

\subsection{Logical error rate results}\label{subsec:logical_results}

We now present logical error per EC cycle for the three architectures under the noise model above.
For the SM-SC, results are obtained from Stim+MWPM simulations.
For the MFEC architectures, we perform Monte Carlo sampling over effective locations using the resource counts from Table~\ref{tab:resources} and Appendix~\ref{app:gatecounts}.
Unless stated otherwise, we consider distances $d\in\{3,5,7,9,11\}$ and use $N_{\mathrm{shots}}=2000$ samples per parameter point.

\subsubsection{Dependence on measurement-induced decoherence}

We first study how performance changes as measurements become slower and noisier compared to coherent operations.
To isolate this effect, we fix the coherence times at their base values
\begin{equation}
    T_1 = T_1^{\mathrm{(base)}}, \quad
    T_2 = T_2^{\mathrm{(base)}},
\end{equation}
and vary $t_{\mathrm{m}}$ to scan the decoherence ratio
\begin{equation}
    R_{\mathrm{decoh}}
    = \frac{p_{\mathrm{decoh}}(t_{\mathrm{m}};T_1,T_2)}
           {p_{\mathrm{decoh}}(t_{\mathrm{idle,op}};T_1,T_2)}
\end{equation}
over a range $R_{\mathrm{decoh}}\in\{20,50,100,200,400,800\}$.

\begin{figure}[htb!]
    \centering
    \includegraphics[width=1.1\textwidth]{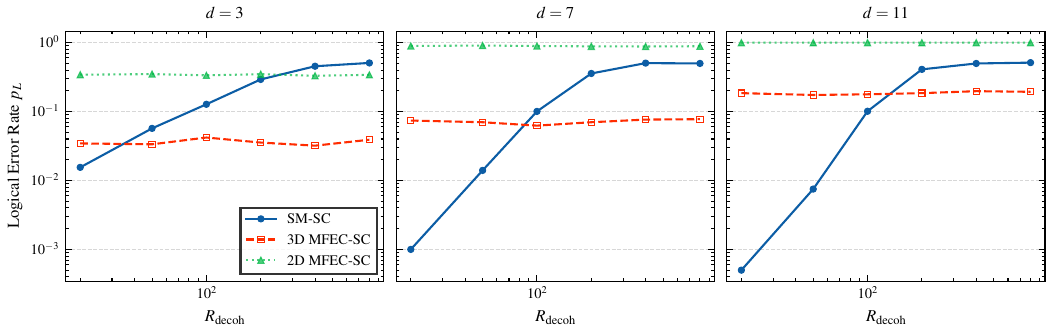}
    \caption{Logical error per EC cycle as a function of the decoherence ratio
    $R_{\mathrm{decoh}} = p_{\mathrm{decoh}}(t_{\mathrm{m}};T_1,T_2)/p_{\mathrm{decoh}}(t_{\mathrm{idle,op}};T_1,T_2)$
    at fixed coherence times $(T_1,T_2)=(200~\mu\mathrm{s},150~\mu\mathrm{s})$, for distances
    $d\in\{3,7,11\}$.
    Solid: standard measurement-based surface code (SM-SC, Stim+MWPM).
    Dashed: proposed 3D MFEC-SC. Dotted: 2D MFEC-SC.
    The SM-SC degrades as measurements become relatively noisier, while the 3D MFEC-SC is essentially insensitive to $R_{\mathrm{decoh}}$.
    The 2D MFEC-SC remains significantly worse due to SWAP overhead.}
    \label{fig:ler_vs_Rdecoh_d3_7_11}
\end{figure}

Fig.~\ref{fig:ler_vs_Rdecoh_d3_7_11} shows the logical error per EC cycle as a function of $R_{\mathrm{decoh}}$ for $d=3,7,11$.
As expected, the SM-SC becomes progressively worse as measurement decoherence grows, because each cycle necessarily spends a fixed fraction of time in noisy measurement windows.
By contrast, the 3D MFEC-SC curves are almost flat: their EC cycles do not contain mid-circuit measurements, so the logical error is essentially independent of $R_{\mathrm{decoh}}$. For the MFEC architectures, the logical error does not depend on the measurement-induced decoherence ratio
$R_{\mathrm{decoh}}$, since the MFEC cycles contain no mid-circuit measurements. In Fig.~\ref{fig:ler_vs_Rdecoh_d3_7_11},
the residual variation of the MFEC curves with $R_{\mathrm{decoh}}$ is a Monte-Carlo sampling artifact; within our model,
$p_L^{\mathrm{MF}}$ is strictly independent of $R_{\mathrm{decoh}}$.

The 2D MFEC-SC remains at much higher error rates for all distances, reflecting the SWAP-induced depth and idling overhead.

\begin{table}[htb]\centering
\caption{Estimated break-even decoherence ratio $R_{\mathrm{decoh}}^{\ast}(d)$ at which the SM-SC and 3D MFEC-SC have comparable logical error per EC cycle, extracted from the data in Fig.~\ref{fig:ler_vs_Rdecoh_d3_7_11}.}
\label{tab:Rcrit}
\renewcommand{\arraystretch}{1.3}          % 행 간격 조금 늘리기
\setlength{\tabcolsep}{16pt}              % 열 간격 늘려서 가로폭 채우기
\begin{tabular*}{\textwidth}{@{\extracolsep{\fill}} ccc}
\toprule
Code distance $d$ &
$R_{\mathrm{decoh}}^{\ast}(d)$ (approx.) &
Regime \\ 
\midrule
$3$  & $\sim 3\times 10^{1}$–$5\times 10^{1}$  & early crossover \\
$5$  & $\sim 5\times 10^{1}$–$8\times 10^{1}$  & moderate measurements \\
$7$  & $\sim 1\times 10^{2}$                   & strongly noisy measurements \\
$9$  & $\sim 1$–$2\times 10^{2}$               & strongly noisy measurements \\
$11$ & $\sim 2\times 10^{2}$                   & very noisy measurements \\
\bottomrule
\end{tabular*}
\end{table}

For each $d$, we define a break-even ratio $R_{\mathrm{decoh}}^{\ast}(d)$ at which the SM-SC and 3D MFEC-SC attain comparable logical error per EC cycle. The extracted values are summarized in Table~\ref{tab:Rcrit}. For small distances ($d=3,5$), the crossover already occurs at relatively modest ratios, $R_{\mathrm{decoh}}^{\ast}\sim 3\times 10^{1}$–$8\times 10^{1}$, i.e., when measurement-induced decoherence exceeds idle decoherence by only a few tens. As $d$ increases, the SM-SC benefits from its shallow depth while the 3D MFEC-SC accumulates more gate and idle errors, so $R_{\mathrm{decoh}}^{\ast}(d)$ gradually shifts upward. Even up to $d=11$, however, the required ratio remains in the range of $\mathcal{O}(10^{2})$, indicating that measurement decoherence need only be roughly two orders of magnitude larger than idle decoherence for the 3D MFEC-SC to become competitive or favorable.

Across the same parameter range, the 2D MFEC-SC never reaches parity with the SM-SC. Its logical error remains substantially higher, especially for larger $d$, underscoring that MFEC combined with planar SWAP-based routing is not a practical solution to the measurement bottleneck.

\subsubsection{Dependence on overall noise scale}

We next vary the overall noise level while keeping the relative decoherence ratio fixed.
To this end, we rescale the coherence times via a noise scale $s$,
\begin{equation}
    T_1 = T_1^{\mathrm{(base)}}/s, \qquad
    T_2 = T_2^{\mathrm{(base)}}/s,
\end{equation}
and, for each $s$, choose $t_{\mathrm{m}}$ such that $R_{\mathrm{decoh}}$ remains constant (here we take $R_{\mathrm{decoh}}=200$ as a representative value).
We then sweep $s$ over a region $s\in[0.5,2]$, corresponding to roughly a factor of four variation in decoherence rates.

\begin{figure}[htb!]
    \centering
    \includegraphics[width=1.1\textwidth]{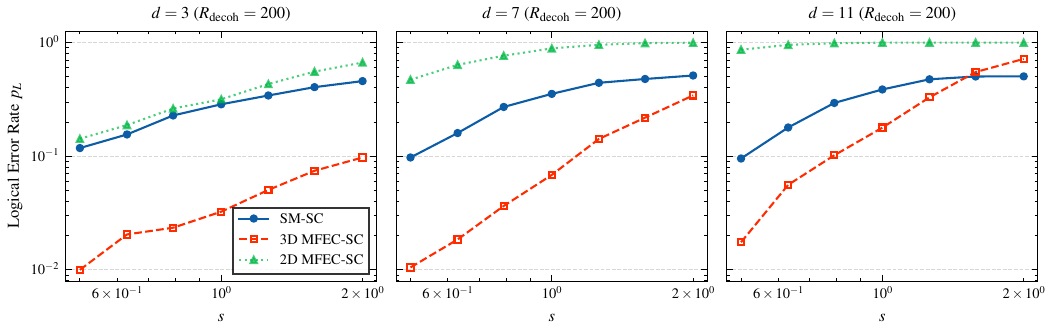}
    \caption{Logical error per EC cycle as a function of the overall noise scale $s$, which rescales $(T_1,T_2)\rightarrow(T_1^{\mathrm{(base)}}/s,T_2^{\mathrm{(base)}}/s)$, at fixed $R_{\mathrm{decoh}}=200$, for $d\in\{3,7,11\}$.
    For small and intermediate distances, the 3D MFEC-SC yields lower logical error than the SM-SC over a wide range of $s$, because it avoids noisy measurement windows.
    At larger distances and very high noise, the deeper MFEC cycles start to lose their advantage.
    The 2D MFEC-SC remains worse across all cases.}
    \label{fig:ler_vs_noise_d3_7_11}
\end{figure}

Fig.~\ref{fig:ler_vs_noise_d3_7_11} shows the resulting logical error rates for $d=3,7,11$.
For small and moderate distances, the 3D MFEC-SC consistently achieves lower logical error than the SM-SC across the entire range of $s$.
The SM-SC suffers from the fixed cost of noisy measurement windows, so its curves flatten or saturate once measurement errors dominate.
In contrast, the 3D MFEC-SC trades measurement noise for additional coherent gates, which remain relatively benign as long as gate errors are in the $10^{-4}$–$10^{-3}$ range.
At larger distances and very high noise levels (large $s$), the deeper MFEC cycles eventually become competitive or slightly worse than the SM-SC, as expected.

\begin{figure}[htb!]
    \centering
    \includegraphics[width=0.6\textwidth]{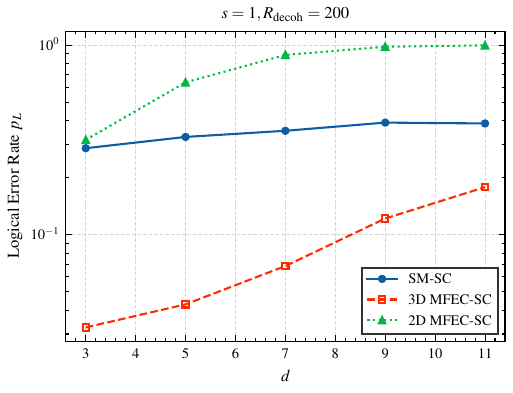}
    \caption{Logical error per EC cycle as a function of code distance $d$ for the three architectures at a fixed working point $(R_{\mathrm{decoh}}=200,s=1)$.
    The 3D MFEC-SC (dashed) increases only mildly with $d$.
    The SM-SC (solid) initially improves with distance but quickly saturates due to measurement-induced errors.
    The 2D MFEC-SC (dotted) exhibits near-unit logical error for $d\ge 7$, highlighting the severe impact of SWAP overhead in planar MFEC implementations.}
    \label{fig:ler_vs_distance}
\end{figure}

Finally, to visualize how logical error scales with distance at a fixed working point, Fig.~\ref{fig:ler_vs_distance} plots $p_L$ as a function of $d$ for $(R_{\mathrm{decoh}}=200,s=1)$.
The SM-SC exhibits an initial decrease with distance followed by saturation near $p_L\sim 0.3$–$0.4$, reflecting the measurement-induced floor.
The 3D MFEC-SC shows a relatively mild increase with distance, remaining competitive or better than the SM-SC for all $d$ in this parameter regime.
The 2D MFEC-SC, on the other hand, already has high logical error at $d=3$ and is essentially non-functional (near-unit error) for $d\ge 7$.

Overall, these results support the architectural message of this work.
In regimes where measurement-induced decoherence is one to two orders of magnitude larger than decoherence during coherent operation idling—a regime that is relevant for many current superconducting devices—the 3D MFEC surface-code architecture can substantially outperform both a standard measurement-based surface code and a 2D MFEC surface code.
The improvement does not come from more aggressive coding or exotic gate gadgets, but from using 3D connectivity as a logical resource to remove SWAP chains and enable constant-depth transversal interactions between logical patches.
A more detailed breakdown of gate counts and scheduling is provided in Appendix~\ref{app:gatecounts}.

%%%%%%%%%%%%%%%%%%%%%%%%%%%%%%%%%%%%%%%%%%%%%%%%%%%%%%%%%%%%%
\section{Conclusion}\label{Conclusion}

We proposed a 3D stacked surface-code architecture for measurement-free fault-tolerant quantum error correction. The core idea is to place three rotated surface-code patches---encoding $\ket{0}_L$, $\ket{\psi}_L$, and $\ket{+}_L$---on separate layers and connect vertically aligned qubits by native couplers. This layout enables transversal CNOT gates between logical patches in constant depth with respect to the code distance $d$, thereby eliminating the $O(d)$ SWAP overhead and associated hook-error mechanisms that arise in purely 2D nearest-neighbor implementations.

On top of this architecture, we constructed a measurement-free QEC protocol that uses only standard Clifford operations and logical auxiliary patches, without resorting to highly complex gate gadgets. Our coarse-grained analysis under a decoherence-based noise model indicates that, in regimes where measurements are significantly slower and noisier than entangling gates, the proposed 3D MFEC surface-code architecture can achieve higher logical fidelity per QEC cycle than both a conventional measurement-based surface code and a 2D MFEC surface code with SWAP-based routing. This supports the view that 3D integration can be used not only for wiring convenience, but as a genuine logical resource for mitigating the measurement bottleneck.

Future work includes full circuit-level simulations for higher-distance codes with realistic decoders, hardware-specific optimization of the 3D layout and vertical couplers, and extending the present framework to lattice surgery, logical CNOT between distant patches, and non-Clifford resources. Taken together, these directions aim to develop 3D MFEC surface-code architectures into a practical route toward high-performance fault-tolerant quantum computing in platforms where slow, error-prone measurements are a dominant limitation.

\backmatter

\bmhead{Acknowledgments}
TBD

\bmhead{Author Contributions}

GSM conceived the proposed method, performed the simulations, derived the analytical equations, and prepared the figures. IKS contributed to the interpretation of the results and to drafting and revising the manuscript. JH contributed to the development of the core ideas and supervised the project.

\bmhead{Data availability}

The datasets generated and/or analysed during the current study are available from the corresponding author on reasonable request.

\bmhead{Conflict of interest}

The authors declare that they have no conflict of interest.

\bmhead{Funding}

Not applicable.

\bmhead{Ethics approval and consent to participate}

Ethical review and approval were not required for this study.

\begin{appendices}

\section{Example MFEC construction for a small code}\label{app:bitflip}

For completeness, we briefly sketch a measurement-free QEC (MFEC) construction for the three-qubit bit-flip code, which serves as a pedagogical example of the abstract structure in Eqs.~\eqref{eq:generic_ext}–\eqref{eq:generic_corr}.

The bit-flip code encodes one logical qubit into three physical qubits with logical basis states
\begin{equation}
    \ket{0}_L = \ket{000},\qquad
    \ket{1}_L = \ket{111},
\end{equation}
and stabilizer generators
\begin{align}
    S_1 &= Z\otimes Z \otimes I,\\
    S_2 &= I\otimes Z \otimes Z,
\end{align}
with $S_3=S_1 S_2 = Z\otimes I \otimes Z$ providing redundant syndrome information.

A measurement-based correction step would measure $S_1$ and $S_2$, decode the two-bit syndrome classically, and apply a conditional $X$ correction to the appropriate physical qubit. In an MFEC construction, we instead introduce three ancilla qubits $a_1,a_2,a_3$ initialized in $\ket{0}^{\otimes 3}$ and proceed coherently:

\begin{enumerate}
    \item \textbf{Syndrome extraction.}  
    Coherently map the stabilizer eigenvalues onto the ancillas by applying controlled-$Z$-type circuits, which can be decomposed into CNOTs and single-qubit gates. For example,
    \begin{align}
        &\mathrm{CNOT}(1\rightarrow a_1),\ \mathrm{CNOT}(2\rightarrow a_1),\\
        &\mathrm{CNOT}(2\rightarrow a_2),\ \mathrm{CNOT}(3\rightarrow a_2),
    \end{align}
    followed by suitable single-qubit rotations on $a_1,a_2$ to map phase information into computational-basis syndromes. The redundant stabilizer $S_3$ can be extracted in a similar fashion onto $a_3$.

    \item \textbf{Coherent correction.}  
    Implement a correction unitary of the form
    \begin{equation}
        U_{\mathrm{corr,bit}}
        = \sum_{s\in\{0,1\}^3} P_s \otimes C_s,
    \end{equation}
    where $P_s$ projects the ancilla register onto syndrome state $\ket{s_1 s_2 s_3}$ and $C_s$ is an $X$ on the appropriate data qubit (or identity) implemented via Toffoli and triple-controlled NOT gates controlled by $(a_1,a_2,a_3)$.

    \item \textbf{Reset.}  
    After the coherent correction, the ancillas are reset to $\ket{0}^{\otimes 3}$ and the cycle can be repeated.
\end{enumerate}

With a suitable choice of $C_s$ and circuit scheduling, it can be shown that any single fault in this MFEC step produces at most one effective $X$ error on the data block at the end of the cycle, which is correctable by the code. Logical failure rates at low physical error probabilities $p$ then scale as $O(p^2)$, consistent with distance $d=3$~\cite{Perlin2023FTMF,Heussen2024MFEC}. The surface-code MFEC constructions in the main text follow the same logical pattern but use a larger set of stabilizers and a logical auxiliary block instead of unencoded ancillas to mediate corrections.

\section{Gate-count and depth estimates for the 3D MFEC surface code}\label{app:gatecounts}

In this appendix we provide explicit gate-count and depth estimates used in Sec.~\ref{sec:analysis} to define effective error-location counts for the 2D and 3D MFEC surface-code architectures. We focus on the distance-$3$ instance as a concrete example and outline how the scaling with $d$ arises.

\subsection{Decomposition of multi-controlled gates}

Our MFEC protocol uses Toffoli (CCNOT) and triple-controlled NOT (C$^3$NOT) gates. To estimate resource overhead we decompose these gates into CNOT and single-qubit gates.

For the Toffoli gate we use the optimal 6-CNOT decomposition of Shende \textit{et al.}, which minimizes the CNOT-count under all-to-all connectivity~\cite{ShendeToffoli},
\begin{equation}
    N_{\mathrm{CNOT}}(\mathrm{Toffoli}) = 6.
\end{equation}
For the triple-controlled NOT we start from an optimized decomposition of the triple-controlled $Z$ (CCCZ) gate~\cite{NakanishiCCCZ}, which achieves
\begin{equation}
    N_{\mathrm{CNOT}}(\mathrm{CCCZ}) = 7.
\end{equation}
Multi-controlled $X$ and $Z$ gates are locally Clifford-equivalent, so we take
\begin{equation}
    N_{\mathrm{CNOT}}(\mathrm{C}^3\mathrm{NOT}) = 7.
\end{equation}
In both cases we conservatively assume that the CNOT-depth is of the same order as the CNOT-count; routing overhead within a surface-code patch is not included at this level, since our focus is on comparing 2D and 3D implementations at the logical level.

\subsection{Distance-3 $X$-error correction round in 3D MFEC-SC}

We now count CNOT locations for the $X$-error correction round in the 3D MFEC surface-code architecture for a single logical qubit encoded in a distance-$3$ rotated surface code. The three stacked patches encode $\ket{0}_L$ (upper layer), $\ket{\psi}_L$ (middle, data layer), and $\ket{+}_L$ (lower layer), with vertical couplers as in Sec.~\ref{subsec:3d_arch_and_cycle}.

The $X$-error correction round couples the data patch $\ket{\psi}_L$ to the $\ket{+}_L$ patch using $Z$-type stabilizers and proceeds in four main stages:

\begin{enumerate}
    \item \textbf{Transversal coupling.}  
    A transversal CNOT is applied from the data patch to the $\ket{+}_L$ patch. For distance $d=3$ there are $9$ data qubits, so
    \begin{equation}
        N_{\mathrm{CNOT}}^{(1)} = 9.
    \end{equation}
    All nine vertical CNOTs act on disjoint pairs and can be executed in a single time step.

    \item \textbf{Syndrome extraction.}  
    $Z$-type stabilizers on the $\ket{+}_L$ patch are coherently mapped onto four unencoded ancillas using 12 CNOTs, arranged in four parallel layers. Thus
    \begin{equation}
        N_{\mathrm{CNOT}}^{(2)} = 12.
    \end{equation}

    \item \textbf{Syndrome remapping.}  
    After resetting a subset of $\ket{+}_L$ qubits, the extracted syndrome bits are remapped from the ancillas onto this compact register using three CNOTs:
    \begin{equation}
        N_{\mathrm{CNOT}}^{(3)} = 3.
    \end{equation}

    \item \textbf{Coherent correction.}  
    The three-qubit syndrome register then controls three C$^3$NOT gates and four Toffoli gates that apply $X$ corrections to selected data qubits. Using the decompositions above,
    \begin{align}
        N_{\mathrm{CNOT}}^{(4)}
        &= 3 N_{\mathrm{CNOT}}(\mathrm{C}^3\mathrm{NOT})
           + 4 N_{\mathrm{CNOT}}(\mathrm{Toffoli}) \nonumber\\
        &= 3\times 7 + 4\times 6 = 45.
    \end{align}
\end{enumerate}

Collecting all contributions, the distance-3 $X$-error correction round uses
\begin{equation}
    N_{\mathrm{CNOT}}^{(X)} = 9 + 12 + 3 + 45 = 69
\end{equation}
two-qubit gates. The corresponding CNOT-depth is dominated by the multi-controlled corrections and is of order 50 time steps in our scheduling.

The $Z$-error correction round, which couples the data patch $\ket{\psi}_L$ to the $\ket{0}_L$ patch using $X$-type stabilizers, has an analogous structure and we estimate
\begin{equation}
    N_{\mathrm{CNOT}}^{(Z)} \approx N_{\mathrm{CNOT}}^{(X)} \approx 69.
\end{equation}

A full EC cycle (one $X$ round plus one $Z$ round) therefore uses
\begin{equation}
    N_{\mathrm{CNOT}}^{\mathrm{cycle}} \approx 138
\end{equation}
two-qubit gates. However, thanks to the 3D stacking, the depth of the cycle is closer to the \emph{maximum} of the two round depths than to their sum: the heaviest parts of the $X$ and $Z$ routines act predominantly within the $\ket{+}_L$ and $\ket{0}_L$ ancilla layers and can overlap in time while the data layer is idle or engaged in transversal couplings. In our estimates we take a representative CNOT-depth per EC cycle of 59 for $d=3$, consistent with this pipelined schedule and with the values reported in Table~\ref{tab:resources}. For $d=5$ the same reasoning leads to a depth scaling as $O(d^2)$ with a coefficient of order 7, as indicated in the table.

\subsection{Effective error-location counts}

For the coarse-grained MFEC error model in Sec.~\ref{subsec:logical_results}, we define effective two-qubit gate and idle-location counts per EC cycle as
\begin{align}
    N_{\mathrm{gate}}^{(3\mathrm{D})} &\approx N_{\mathrm{CNOT}}^{\mathrm{cycle}},\\
    N_{\mathrm{idle}}^{(3\mathrm{D})} &\approx \alpha\, N_{\mathrm{gate}}^{(3\mathrm{D})},
\end{align}
with $\alpha$ a constant of order 3–4 that accounts for the fact that, in each gate layer, several qubits are idle while others participate in operations. For 2D MFEC-SC, we scale $N_{\mathrm{gate}}$ and $N_{\mathrm{idle}}$ by the ratio of CNOT-depths between 2D and 3D (Table~\ref{tab:resources}), reflecting the SWAP-induced increase in depth and idling.

These effective location counts, together with the noise probabilities in Table~\ref{tab:params}, define the binomial distributions used to sample the number of faulty locations in each MFEC EC cycle in Sec.~\ref{subsec:logical_results}. Although coarse-grained, this model faithfully captures the leading effects of depth and connectivity on logical performance and allows us to compare 2D and 3D MFEC architectures under the same physical noise assumptions.

\section{2D MFEC circuits and SWAP-based transversal routing}\label{app:2d_mfec_details}

\begin{figure}[htb!]
    \centering
    \includegraphics[width=0.75\textwidth]{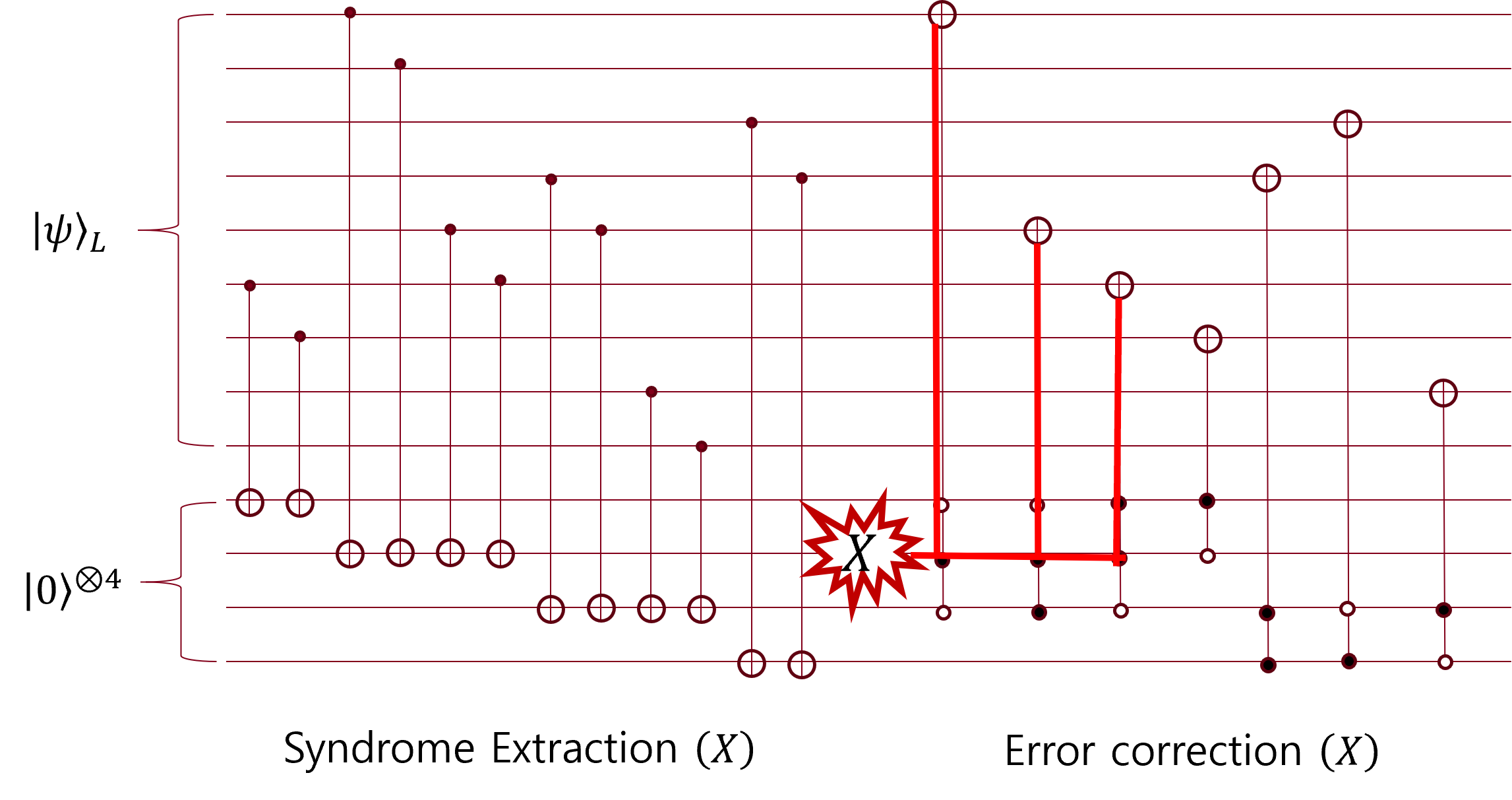}
    \caption{Naive MFEC surface-code correction circuit (distance-$3$ example). A single fault on a syndrome ancilla can propagate into an uncorrectable multi-qubit error on the data block~\cite{Aliferis08}.}
    \label{app:fig:circuitX}
\end{figure}

\begin{figure}[htb!]
    \centering
    \includegraphics[width=0.65\textwidth]{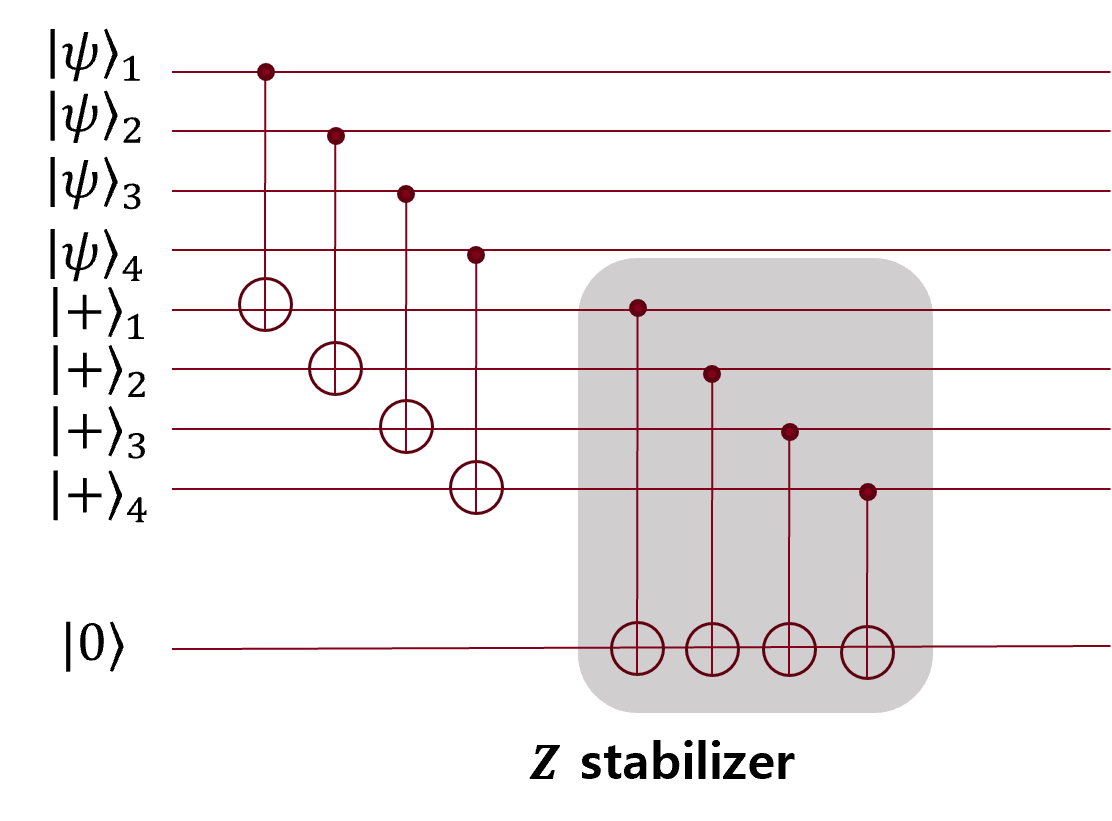}
    \caption{Fault-tolerant MFEC idea in 2D: introduce a logical auxiliary patch and use transversal coupling to mediate coherent feedback, rather than letting unencoded syndrome ancillas directly control multi-target physical corrections~\cite{Heussen2024MFEC,Veroni2024OptimizedMFEC}.}
    \label{app:fig:ft}
\end{figure}
This appendix collects circuit-level details for implementing MFEC surface-code steps on a strictly planar 2D nearest-neighbor (NN) architecture. These details are omitted from the main text to keep the focus on the proposed 3D architecture, but they provide a concrete illustration of why planar routing creates an $O(d)$ depth overhead for transversal coupling and why this overhead can become performance-limiting~\cite{Aliferis08,Veroni2024OptimizedMFEC,Heussen2024MFEC}.

Fig.~\ref{app:fig:circuitX} shows a representative \emph{naive} measurement-free construction for correcting $X$ errors via $Z$-type stabilizers in a distance-$3$ rotated surface code~\cite{Veroni2024OptimizedMFEC}. Although syndrome information can be extracted coherently, directly using unencoded syndrome ancillas to drive multi-target corrections is not fault-tolerant: a single fault on a control ancilla can propagate into a correlated multi-qubit error on the data block, violating the standard distance-$d$ fault-tolerance condition~\cite{Aliferis08}. A standard remedy is to introduce an additional logical auxiliary patch (e.g., prepared in $\ket{+}_L$ for $X$-error correction), and to mediate coherent feedback through transversal interactions between the data and auxiliary blocks, so that any single physical fault can be arranged to produce at most one effective error per code block at the end of the EC step~\cite{Perlin2023FTMF,Heussen2024MFEC,Veroni2024OptimizedMFEC}. Fig.~\ref{app:fig:ft} illustrates this fault-tolerant structure at a schematic level.

\begin{figure*}[htb!]
    \centering
    \includegraphics[width=\textwidth]{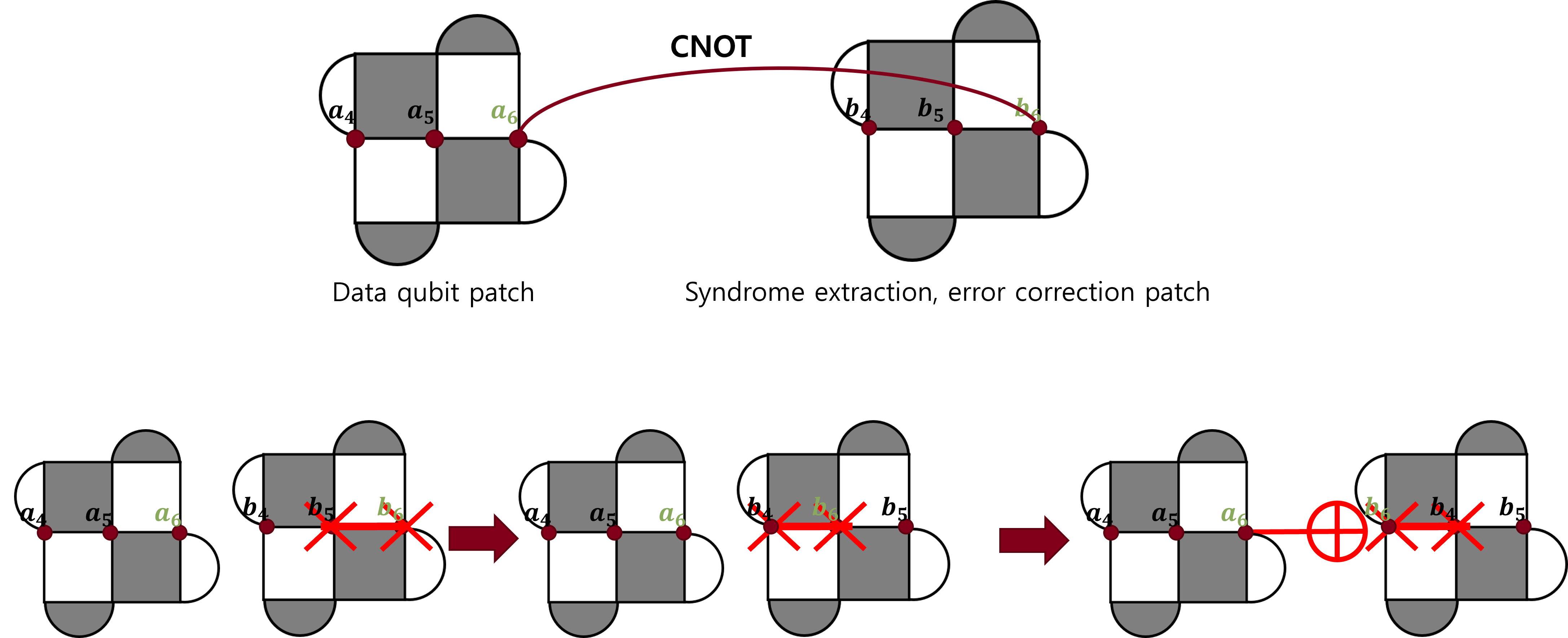}
    \caption{Example schedule for a transversal CNOT between two distance-$3$ surface-code patches on a 2D NN layout. The required SWAP routing produces a depth overhead that scales as $O(d)$ for distance $d$.}
    \label{app:fig:2d_seq}
\end{figure*}
In a planar 2D NN layout, however, corresponding physical qubits in neighboring logical patches are separated by $O(d)$ lattice steps. Implementing a transversal CNOT layer therefore requires SWAP routing of length $O(d)$ per qubit pair, which increases depth, gate count, and idling time~\cite{Aliferis08}. Fig.~\ref{app:fig:swap_circuit} depicts the conceptual idea of SWAP-based remote coupling, and Fig.~\ref{app:fig:2d_seq} shows an example schedule for a distance-$3$ transversal CNOT between two surface-code patches. Even when the logical MFEC design is fault-tolerant~\cite{Perlin2023FTMF,Heussen2024MFEC}, this routing overhead can dominate the effective number of fault locations per EC cycle due to (i) the $O(d)$ depth increase per transversal coupling layer and (ii) additional idling during routing. This is the primary reason that planar 2D MFEC surface-code implementations can remain uncompetitive relative to standard measurement-based surface-code cycles in realistic parameter regimes, whereas the proposed 3D stacked architecture restores transversal coupling without SWAP routing and recovers the intended MFEC advantage in the slow/noisy measurement regime~\cite{suppressing_2023,Graham23,Lis23}.

\end{appendices}

\bibliography{sn-bibliography}
\end{document}